\documentclass{sig-alternate-10pt}

\usepackage{amsmath,amssymb,array,color,colortbl,graphicx,multirow}
\usepackage{microtype}
\usepackage{comment}

\usepackage[bookmarks=false]{hyperref}

\usepackage{multirow}

\usepackage{algorithm2e}
\usepackage{comment}

\usepackage{framed,color}

\definecolor{mygreen}{rgb}{0,0.5,0}

\newcommand{\sys}{PRI\xspace}

\begin{document}

\title{\sys: Privacy Preserving Inspection of\\Encrypted Network Traffic}



\author{Anonymous}

\author{
Liron Schiff$^1$ \quad Stefan Schmid$^{2}$\\
{\small~$^1$ Tel Aviv University, Israel \quad $^2$ Aalborg University, Denmark}
}

\date{}

\maketitle \thispagestyle{empty}

\sloppy

\begin{abstract}
Traffic inspection is a fundamental building block of many 
security solutions today. For example,  
to prevent the leakage or exfiltration of confidential insider information,
as well as to block malicious traffic from entering the network,
most enterprises today operate intrusion detection and 
prevention systems that inspect traffic. 
However, the state-of-the-art inspection systems 
do not reflect well the interests of the different
involved autonomous roles.
For example, employees in an enterprise, or a company
outsourcing its network management to a specialized third party,
may require that
their traffic remains confidential, even from the 
system administrator. 
Moreover, the rules used by the intrusion detection
system, or more generally the configuration of an 
online or offline anomaly detection engine,
may
be provided by a third party, e.g.,
a security research firm, and can hence constitute
a critical business asset which should 
be kept confidential.
Today, it is often believed that 
accounting for these additional
requirements is impossible, as they 
contradict efficiency and effectiveness. 
We in this paper explore a novel approach,
called
Privacy Preserving Inspection~(\sys),
which provides a solution to this problem,
by preserving  
privacy of traffic inspection
and confidentiality of inspection rules and configurations,
and e.g., also supports the flexible installation of 
additional Data Leak Prevention~(DLP) rules 
specific to the company.
\end{abstract}

\section{Introduction}\label{sec:intro}

The Internet has become a critical and indispensable infrastructure 
for many organizations. At the same time, the Internet constitutes a security threat.
For example, web-based services, such as email, 
are indispensable for communicating with others either within 
or outside
of an organization, but introduce the risk of data exfiltration. 

Intrusion Detection Systems~(IDS) as well as Intrusion
Prevention Systems~(IPS) are frequently used 
today to defend networks~(or specific servers) from cyber attacks~\cite{bro,snort}.
In particular, such systems
can prevent exfiltration of confidential insider information
by blocking accidental or intentional leakage, e.g., by searching for document confidentiality watermarks in the
data transferred out of an enterprise network.
Such systems are also vital to control inbound traffic, and 
e.g., to
detect if packets from a compromised
sender contain an attack, 
employ parental
filtering to prevent children from accessing adult material,
etc.~\cite{blindbox}
Indeed, many cyber attacks today are carried out remotely, exploiting 
vulnerabilities in network components or applications, or tempting naive users 
to download malware and install them on their PCs. 

To provide such functionality, these systems rely on
traffic inspection: they allow the definition of
configuration rules which 
define known attack patterns, indicators for attacks, or traffic anomalies,
and which are matched against the packet header \emph{and payload}. 
If rules are matched, alerts are generated, and/or, in case of 
prevention systems or firewalls, packets are
dropped.
The corresponding rules can either be distributed by a
local support team, by a 
third party~(for instance a security research firm), by the network 
operator, or a combination thereof: 
third party provided rules can be complemented with 
organization-specific asset leakage indicators,
coming from Data
Leakage Prevention~(DLP) systems.

While today's intrusion detection and prevention systems typically perform 
well for unencrypted traffic, 
they struggle with encrypted traffic, resulting in false negatives or 
poor performance. As a workaround, in practice today, the secure and 
encrypted channel from or to the Internet is often terminated at a \emph{proxy},
which essentially mounts some kind of ``man-in-the-middle-attack''. 
While this solution ensures an effective detection and prevention, 
it comes at the price that the privacy of user traffic~(e.g., emails) is undermined. 
In fact, even in the case where communication was already unencrypted
anyway, performing deep packet inspection~(not for the purpose 
of forwarding) can be seen as privacy violation. 
Indeed, users and clients have criticized this approach
and expressed worry, e.g., that the private
logged data is given to marketers~\cite{blindbox,unhappy1,unhappy2}.

Privacy-preserving intrusion detection may not only be desirable and relevant in the
context of enterprise networks, but is also gaining in importance in the
light of today's trend to outsource the network management,
including security aspects, to third parties~\cite{someone}.
For example, the management of third-party networks can be a lucrative
business for Internet Service Providers~(ISPs). At the same time, for customers
running security critical businesses~(for example banks), it is important
that the privacy of traffic be preserved. 

We in this paper however observe another confidentiality issue of today's solutions:
it concerns the confidentiality of the inspection
logic itself.  
For example, the development and maintenance of
effective intrusion detection rules is challenging, and
especially
small enterprises do not have the expertise and time to define
the most effective rules and constantly follow the news.
This constitutes a business opportunity for third parties: a company specialized into security
research can take over the responsibility to define and maintain a
good rule set.
However, such a business model also introduces new requirements.
In particular, a third party company may not be willing to share
its rules, or more generally \emph{configurations} of (online or offline) 
anomaly detection systems,
with the customer: these rules and configurations are an intellectual property 
which constitute an essential
asset of the business model.

At first sight, it may appear that the requirements are contradicting:
First, it seems unavoidable that an intrusion detection or prevention system, 
which for efficiency and effectiveness reasons needs to inspect the traffic
in an unencrypted form, may leak information about the user traffic.
Second, it also seems unavoidable that a system administrator operating
a system based on the rules of a third party company, can see the rules.

\subsection{Contribution}

We identify the different autonomous roles involved in 
a traffic inspection system, including intrusion detection/prevention 
systems, but also more sophisticated online or offline anomaly
detection systems as they may for example be required 
to deal with insider threats.
We then explore the feasibility of providing a system
which meets these requirements, respecting the 
autonomy of the different involved stakeholders,
without introducing new threats coming, e.g., from insiders.

In particular, we present a Privacy-Preserving
Intrusion detection/prevention system, short \sys (\emph{fruit} in Hebrew),   
which decouples the different roles,
and hence significantly reduces the required trust assumptions.
\sys leverages the hardware protection of
architectures like Intel SGX~\cite{white} to defend against 
insiders or system administrators aiming
to break the confidentiality. 
A distinguishing feature of \sys
is the simple and cheap deployment: 
a single trusted hardware component
is sufficient. On the user side,
only a simple software update is required. 

\newpage

\subsection{Paper Organization}

The remainder of this paper is organized as follows.
In Section~\ref{sec:model}, we present a model which
identifies the different roles/stakeholders and
the resulting requirements for the traffic inspection 
system. After providing the necessary background~(Section~\ref{sec:back}),
based on these requirements, Section~\ref{sec:ppi}
describes our proposed architecture \sys which meets these requirements.
We discuss use cases in Section~\ref{sec:use-cases} and review related work in Section~\ref{sec:discussion},
and we conclude our contribution in Section~\ref{sec:conclusion}.

\section{Roles and Requirements}\label{sec:model}

In the following, we first
present the different roles
and discuss their objectives.
Based on this model, we then derive
security goals.
Note that for ease of presentation,
in the following, we will mostly focus
on intrusion \emph{detection}.
However,
our approach can 
easily be generalized to \emph{prevention} or \emph{configuration-based} systems, 
as we will elaborate more later in this paper.

We in this paper distinguish between
the following roles:
\begin{itemize}
\item \textbf{The administrators:} The administrators 
are responsible for ensuring the availability and security of the network.
In particular, they want to prevent the leakage of sensitive insider information, 
and also prevent malicious traffic entering the network from the outside.
Besides relying on up-to-date security rules
possibly provided by an external company, 
administrators may also want to be able to add 
their own Data
Leak Prevention~(DLP) rules, specific to their
organization.

\item \textbf{The users:} 
The term user will generally refer to the communication endpoints.
In particular, we will usually assume that one endpoint is inside 
an enterprise~(an \emph{insider}) while the other is outside~(an 
\emph{outsider}); however, also traffic between two insiders can be subject to inspection. 
We assume that while users profit from a secure environment and 
the detection of undesired inbound and outbound traffic, 
 they also desire a high 
communication performance as well as confidentiality of their traffic. 

\item \textbf{The rules (configuration) provider:} The rules 
for the intrusion detection system
(or more generally the configurations for a traffic inspection system based on 
some open-source logic/engine)
may be provided by an external security company
specialized into developing and maintaining high-quality rules.
The rules provider may desire that its rules remain confidential.
\end{itemize}

Given our roles and their objectives, 
we identify the following requirements:
\begin{enumerate}
\item[R1] \textbf{Efficient and effective inspection:}
The inspection system must 
ensure to the administrators that relevant events and attacks will be detected
successfully and fast.

\item[R2] \textbf{Privacy-preserving traffic inspection:} 
We want to ensure to the users that neither the network administrator nor the rule/configuration provider
should be able to see the traffic. 

\item[R3] \textbf{Confidentiality of rules:} 
We want to ensure that the rules are kept confidential
and are not leaked to the other roles.
\end{enumerate}

This paper is motivated by the question whether it is possible
to design an architecture which meets the different goals of
the different roles, maintaining their autonomy. 
Indeed, designing such a system seems challenging:
\begin{itemize}
\item Today's proxy solutions do not meet the
requirements R2 and R3: 
a proxy
server can be exploited by the administrators to learn about the 
unencrypted traffic. 
Rather, we want to develop a solution which does not allow the administrators
to configure the inspection system in a way that allows them to learn details about 
the traffic which are not security relevant. 
\item
Networks usually operate 
at very high rates: 
cryptographic schemes based on
fully homomorphic or functional encryption~\cite{func-encrypt-1,homomorphic1,
func-encrypt-2}
are slow and can decrease network rates by
many orders of magnitude~\cite{aes-eval}, violating requirement R1.
\end{itemize}

At a first
glance, the requirements seem to contradict: a system 
may hardly be able to efficiently and effectively inspect
traffic without introducing opportunities to the administrator
to see the traffic and used rules.
However, as we will see in the following,
there do exist solutions to satisfy 
these seemingly conflicting
properties.

\section{Background}\label{sec:back}

Before presenting our solution,
we provide some background
which is necessary to undertand our solution. In particular, 
we revisit IDS systems and give an introduction to 
SGX.




\subsection{Traffic Inspection Systems}

Almost all cyber security breaches involve transmissions 
of traffic over a network. 
The standard approach to secure transmissions is to
inspect traffic, checking whether the traffic carries the attack 
(e.g., a malware) or its outcome~(e.g., a stolen digital asset).
For this purpose, organizations deploy packet inspection systems in 
their networks, configuring them with known attack indicators, in the case of 
Intrusion Detection/Prevention Systems~(IDS/IPS), and with~(possibly organization-specific) 
asset leakage indicators in the case of Data Leakage Prevention~(DLP) systems.

All inspection systems essentially search for the configured indicators inside the traffic, where indicators can be based on exact match strings, 
regular expressions, statistical properties, and more. Inspection of traffic in these systems usually includes the inspection of the \emph{packet payload}, i.e., accessing the application layer data. 

A simpler type of inspection that considers only the packet headers~(up to the transport layer) is often performed by firewalls. 
A firewall is a network security system that monitors and controls the incoming and outgoing network traffic and may also include IDS/IPS capabilities. Firewalls can be considered  as a barrier between a trusted, secure internal network and another outside network, such as the Internet, thereby mitigating attacks in the early stage of penetration. Firewalls can also be host based, operating on and defending a single machine. 

Inspection systems are also used to detect insider threats~\cite{insi-3}, 
for example by analyzing confidential
documents in private communications, 
possibly enhanced with watermarking techniques~\cite{encryptedweb}.

Web security today is usually realized with HTTPS,
which relies on the
Transport Layer Security~(TLS)  a.k.a.~Secure Sockets Layer~(SSL)
protocol. TLS/SSL 
provides
confidentiality, integrity and authentication of data in transit.
The protocol offers 
encryption, hash functions or message digests, and digital signatures.

Encrypted traffic constitutes a great challenge to packet inspection systems as it hides the payload content from anyone but the session endpoints. 
There are two common 
solutions to this challenge. 
The first is to extend host based firewalls to gain access to the data \emph{before being encrypted}: this can be implemented through integration with the host operating system. 
We will refer to this solution as the \emph{client-side firewall approach}.
The second solution, usually deployed by large firms, is to operate a middlebox that serves as a \emph{proxy} between any internal user PC and any external web server. The implementation of such a solution requires to configure~(trick) the user PCs to identify the proxy as the server and establish an encrypted session with it rather than with the server.

However, both the distributed firewall as well as the proxy architecture 
expose the user data to $3$rd parties, 
namely the firewall producer, the proxy vendor or the network operator, possibly violating the user privacy. In addition, the inspection rules, which are 
the intellectual property of security researchers or firms, might be extracted at the host or the proxy, undermining their effectiveness and profit.

\subsection{SGX}

Intel~\emph{Software Guard Extensions 
(Intel~SGX)}~\cite{sgxr3,sgxr1,sgxr2,white}
are new CPU instructions which allow 
applications to manage private regions of code and data. 
That is, using SGX, 
an application can run in a protected
environment, the so-called \emph{enclave},
secure from malware
or the  
inspection by the computer administrators.

There is no need to encrypt the protected portion of an application 
for distribution. Before the enclave is built, the 
enclave code and data is free for inspection and analysis. 
When the protected portion is loaded into an enclave,
its 
code and data is measured. 
An application can prove its 
identity to a remote party (a procedure called attestation) 
and be securely provisioned 
with keys and credentials. The application can also request an 
enclave and platform specific key that it can use to protect 
keys and data that it wishes to store outside the enclave.
In addition to the security properties, the enclave environment 
offers scalability and performance associated with the execution 
on the main CPU of a platform.

\begin{figure}[t!]
\centering
\includegraphics[width=0.99\columnwidth]{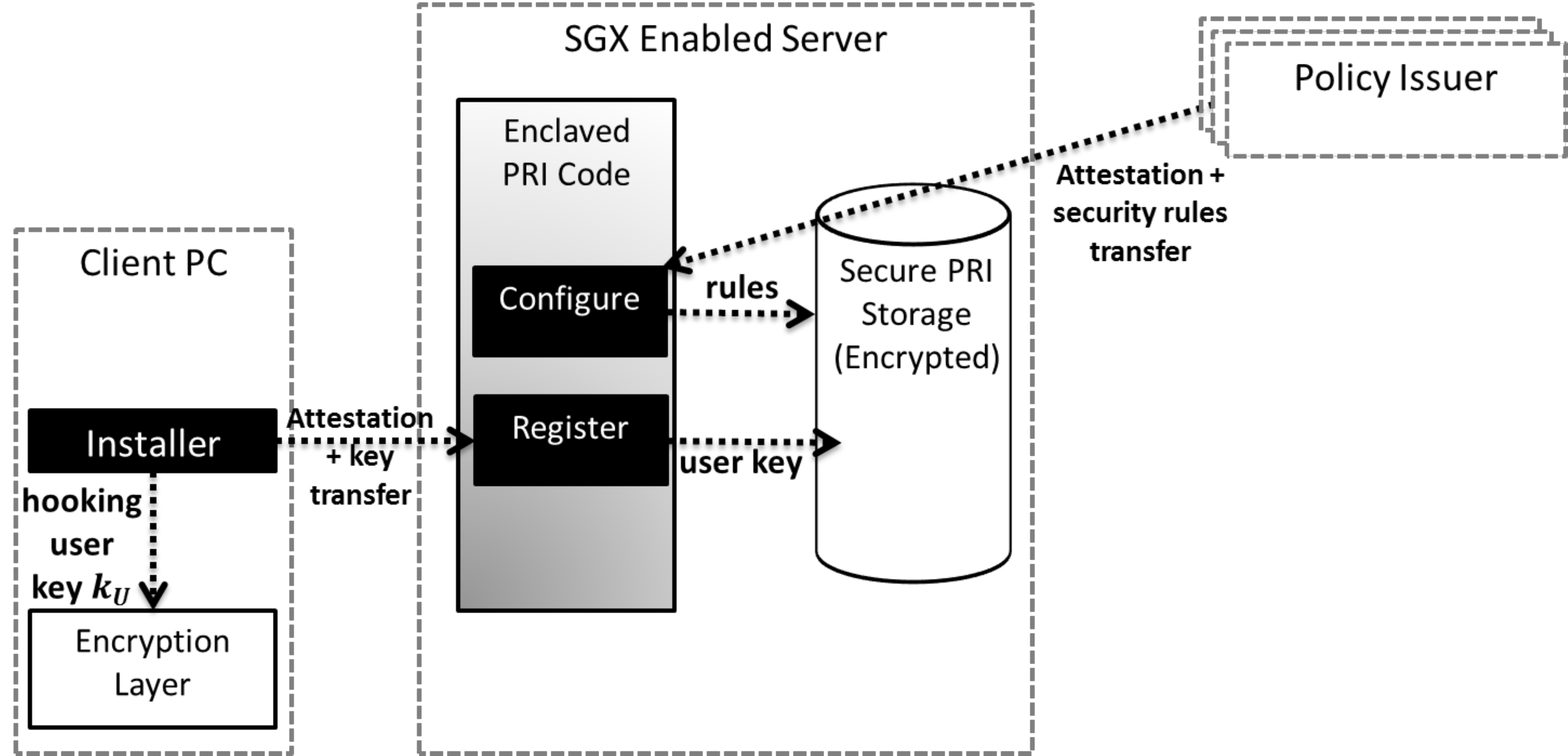}
\caption{
Setup of the PRI system components.
The \sys process is configured to inspect user traffic according to security rules inserted by the policy issuer. The process is secured and ensures confidentiality of traffic and rules. Keys and rules are stored safely in the \sys storage.}\label{fig:fig1}
\end{figure}

\section{Privacy Preserving Inspection}\label{sec:ppi}

This section shows that the design goals and requirements
derived above can actually be met, assuming the availability
of a trusted hardware. In particular, this section 
presents our Privacy Preserving Inspection 
architecture, short \sys. \sys 
allows the inspection, and if considered harmful, possibly prevention,
of encrypted traffic while guaranteeing that no information about the traffic is 
leaked from the inspecting device, and that inspection rules are not revealed to the 
operator of the device nor the traffic generator.

In a nutshell, 
the idea underlying \sys is to decouple
and separate the different roles~(e.g., users, administrators, security company) 
by defining an interface between them and supporting verification.
In order to achieve this deoupling, \sys relies on a single device
(such as Intel SGX or ARM TEE)
which decrypts traffic 
from the users, applies the desired security rules, and when needed 
raises alerts or drops the traffic, without leaking information about the user 
traffic nor the applied rules, to the roles which are not allowed to have this information. 
 Interestingly, a single
trusted box is sufficient for \sys: no hardware modifications are required
on the user side.
Moreover, an important aspect of \sys
is that 
the user side agent and the enclaved system can be 
\emph{open-source}:
thus, the users and the research community can and should verify its privacy preserving property. 
The attestation allows the agent to verify that it  
communicating with the verified system code.

\subsection{Setting Up \sys}

Let us first discuss the setup of \sys~(cf Figure~\ref{fig:fig1}). 
We describe it for the enterprise intrusion detection scenario, but it is easy to adapt 
it to ISP outsourcing scenarios. 
The \sys system consists of the user's communication device, e.g., a personal computer~(PC)~(or smartphone, laptop, etc.) and the \sys server which need to be set up as follows. 
On the user side, we simply need to install 
a \sys agent at the encryption layer of the OS or specific application~(e.g., web browser). The \sys agent is configured with a user key, $k_U$, which is securely transmitted~(through attestation) to the enclaved \sys system.
In addition the agent is installed in a way that allow it to gain the key, $k_S$, of every encrypted session $S$, that uses the encryption layer enabling it to securely send it to the \sys system.

The setup also involves configuring the \sys server with the rules~(also called the policy) 
that need to be checked against the traffic. Rules can be configured by 
multiple policy issuers, each of them can connect securely to the \sys process, 
validating its secure execution~(attestation) and sending the rules. 
As a result, the \sys process stores the rules in the secured storage, 
to be used when a user traffic needs to be inspected.

\subsection{Operating \sys}

\begin{figure}[t!]
\centering
\includegraphics[width=0.99\columnwidth]{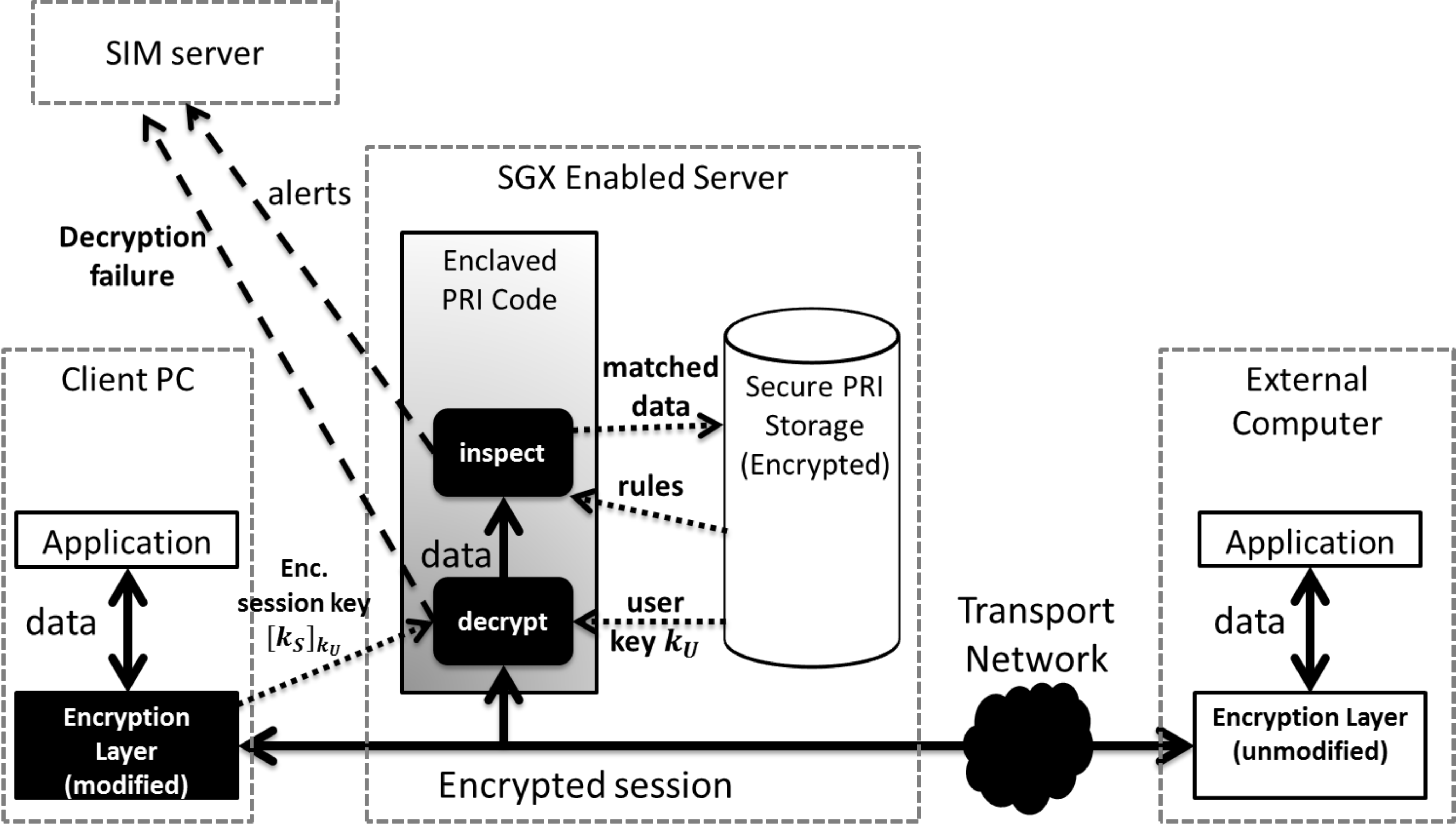}
\caption{The \sys system inpects user session using the key sent by the agent. Matches are saved in a secured storage and alerts are sent to the corporate SIM 
server about detected anomalies.}\label{fig:fig2}
\end{figure}

Once the system has been set up, the registered users' traffic can be inspected. 
Inspection is carried in the following steps, see also Figure~\ref{fig:fig2}.:
\begin{enumerate}
\item 
An encrypted session is established between the applications of the user PC and an external ~(unregistered) or internal PC.
\item The user PC connects to the \sys process~(at the \sys server), and sends to it the session key encripted by the shared user key, i.e., $[k_S]_{k_U}$.
\item The session traffic is duplicated and processed by the \sys process.
\item The session traffic is decrypted, using the session key.
\item  The session data~(clear text) is inspected with the rules configured in the system.
\item  In case part of the data matches a rule, this part is securely stored in the 
\sys storage and an alert~(containing the rule identifier) is reported the corporate 
Security Information Management~(SIM) server.  
\end{enumerate}

Note that in case of misconfiguration, or malicious activity in the user 
PC that results in decryption failure at the \sys system, a special alert is reported. 

\begin{figure}[t!]
\centering
\includegraphics[width=0.99\columnwidth]{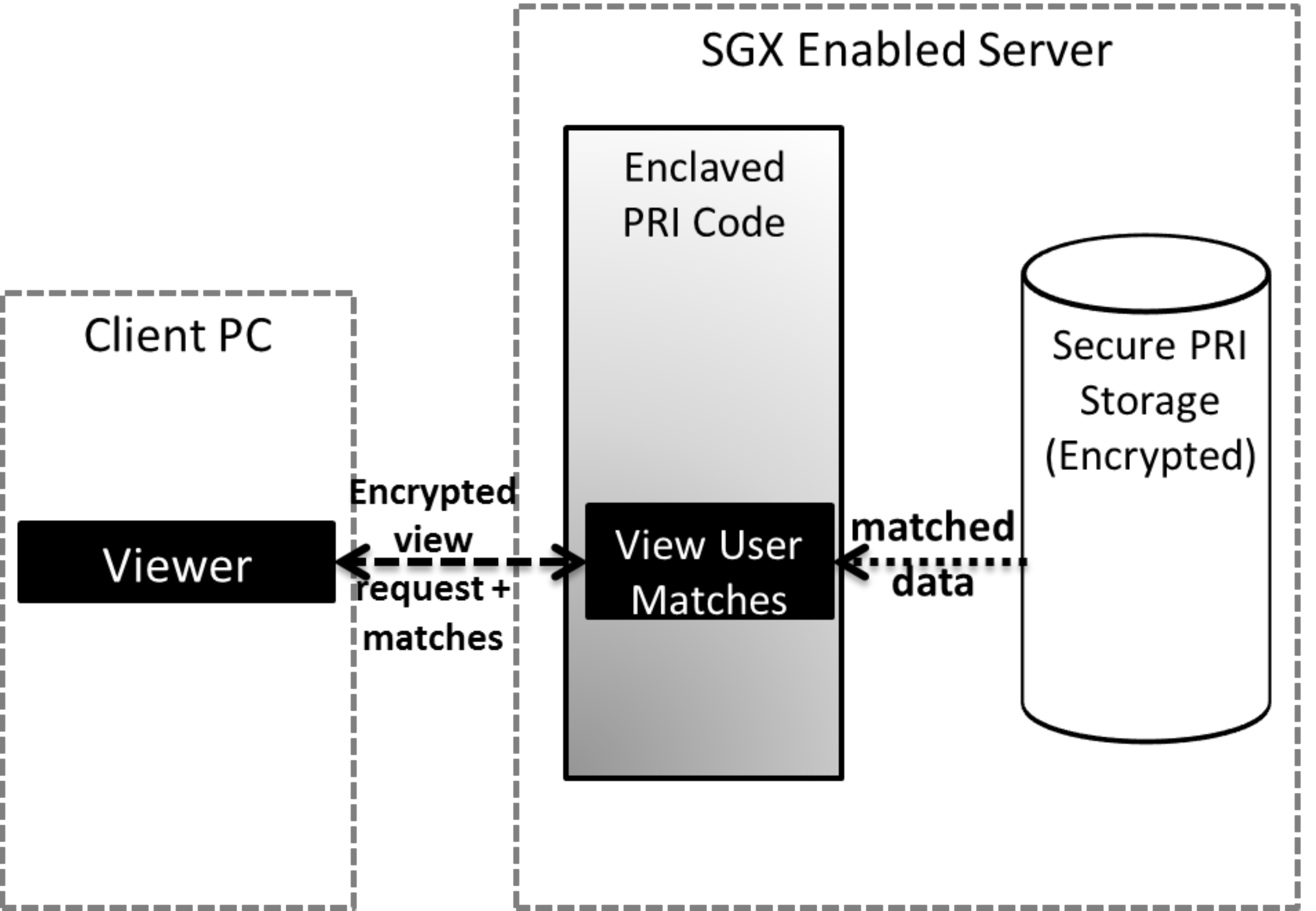}
\caption{The \sys system includes a special viewer app which allows users to verify that matching rules are not abused to violate confidentiality of their traffic.}\label{fig:fig3}
\end{figure}

There exists another more subtle threat for the confidentiality of the traffic: 
even if the traffic cannot be inspected,
the reported matches over traffic 
leak information. 
In order to prevent an attacker from abusing the security rules to 
learn about 
private user data~(e.g., matching every possible byte or word), 
the \sys system allows each user to request the parts of his or her traffic 
which have been matched by the inspecting device, and thus to learn about and 
detect the use of abnormal matching rules. To ensure this, our system 
uses a special viewer application which connects to the \sys process, 
authenticates as the user and issues the request. As a result, the 
\sys process retrieves the user's matches from the secured storage 
and sends them to the viewer app~(cf~Figure~\ref{fig:fig3}.).

In addition, the \sys system can securely 
inspect static~(e.g., likelihood of match for 
common words) and dynamic~(e.g., average number of hits per traffic byte) 
properties of the rules, 
thereby detecting abnormal rule sets.
Also note that inferring traffic content from matches is practically 
difficult in the prevention operation mode of the \sys system, 
since the user session is automatically dropped as a response to the first match.

\begin{figure}[t!]
\centering
\includegraphics[width=0.99\columnwidth]{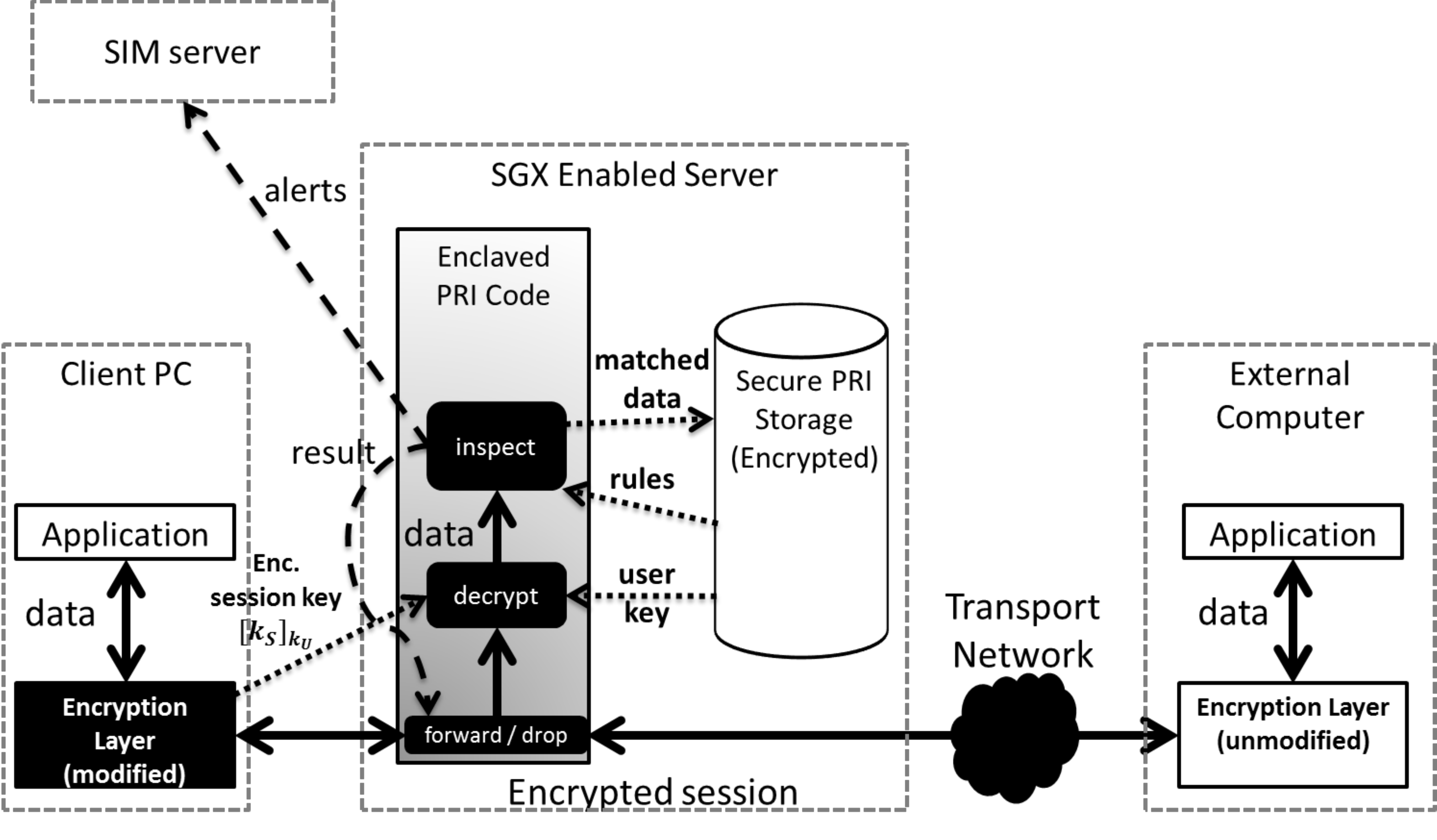}
\caption{The \sys system can also be used for prevention, not only detection. In this case, depending on the result of the inspection, it is decided whether traffic can be forwarded or should be dropped.}\label{fig:fig4}
\end{figure}

The \sys system not only supports intrusion detection but also intrusion prevention,
see Figure~\ref{fig:fig4}: 
The main difference from detection, is that for prevention, 
the outcome of the inspection determines whether the session traffic is 
forwarded through the \sys system, or dropped~(in case of a match).

\subsection{A Note on Implementation}

We started
experimenting with an emulated software version of the SGX 
framework.
However, as the market introduction of the actual 
secure hardware is delayed, our prototype 
is simple and still contains untrusted parts.
Nevertheless, in the following, we report on some preliminary insights.

On the user side, we need to extract encrypted session 
keys and securely transfer them to the \sys server. This can be 
performed in multiple ways.
For example, 
at the application level, applications can report their 
session keys~(as browsers support~\cite{white}) and a \sys agent then 
sends them to the \sys server. Alternatively, applications 
can be extended~(e.g., using a browser plugin) to directly send their key 
to the \sys server.
A third approach is 
to modify 
encryption services offered 
by the operating system, to send the used keys to the \sys server.

\begin{table*}[t]
  \centering
\begin{footnotesize}
\begin{tabular}{cc|c|c|c|c}
\cline{3-5}
& & \multicolumn{3}{ c| }{\textbf{Use Case}}& \\ 
\cline{3-5}
& & \textbf{Enterprise Security and Insiders} & \textbf{Outsourced Security} & \textbf{Anti-Terror Intelligence} \\ \cline{1-5}
\multicolumn{1}{ |c  }{\multirow{3}{*}{\textbf{Entities}} } &
\multicolumn{1}{ |c| }{Clients} & employees & enterprise & civilians / web hosts &     \\ \cline{2-5}
\multicolumn{1}{ |c  }{}                        &
\multicolumn{1}{ |c| }{\sys Operators} &admins & external  & ISPs  &     \\ \cline{2-5}
\multicolumn{1}{ |c  }{}                        &
\multicolumn{1}{ |c| }{Rule Providers} &admins + external & external  & governments  &     \\ \cline{1-5}
\end{tabular}
\end{footnotesize}
\caption{Comparison of \sys use cases.}
\label{tab:tab2}
\end{table*}

On the server side, an inspection program needs to be 
executed in an enclaved environment, using the reported 
user session keys. We argue that adjusting existing IDS systems 
to our needs requires minor code modifications: the computations 
can be performed in user mode 
and the memory consumption can be optimized, 
which should make it easy to enclave them. 
Today there exist several good open-source IDSs such as 
Snort~\cite{snort}, Bro~\cite{bro} and Suricata~\cite{suricata}, 
which we suggest to employ.

Given a selected existing open-source IDS system, 
we propose the following high level design concepts for the \sys server application:
(1) The decryption of traffic can be handled by a 
new enclaved module that pushes the plain text outcome 
to the enclaved and modified IDS system. 
(2) The enclaved IDS system is reduced to only handle protocols 
that are known to be encrypted~(e.g., HTML), or to inspect 
rules that need to be hidden, thereby reducing 
the memory footprint which is limited in the enclaved environment.
(3) In parallel to the enclaved IDS, a standard~(not enclaved) IDS 
is operated for non-encrypted traffic and with open-source rules, 
utilizing the standard less restricted environment.

In our prototype, we attend to the problem of 
human based data leakage, 
and focus
on the decryption  of common webmail services. 
We noticed that this traffic is mainly composed of 
(encrypted) HTTP/2~\cite{http2} traffic which uses HPACK compression. 
This protocol is not supported by the IDSs we examined, and therefore requires 
special adapters.

In general, the code of both the client and server side of the \sys system should be 
fully open, 
allowing the research community to examine them and verify their privacy 
preservation properties. Combining the verification of the code 
with the attestation of the server by the client 
constitutes the basis of the system trust.

\section{Use Cases}\label{sec:use-cases}

We believe that the ideas underlying \sys can be interesting in
several contexts and beyond the standard enterprise network 
security scenario we discussed in this paper so far.
In the following, three examples are discussed in more detail.
A comparison is given in Table~\ref{tab:tab2}.

\subsection{Enterprise Security and Insider Threats}

Email and cloud services (e.g., Google Docs, storage, etc.) are indispensable but introduce
a risk for many organizations today. In particular, such 
services can be used by a malicious
insider to steal intellectual property or sensitive company information. 
Indeed, insider attacks pose some unique challenges for security administrators.
In general, it can be challenging to configure an intrusion detection
system and define a good rule set to detect internal attacks: 
different users should have access to and use many different services, servers, 
and systems for their work. 
However, over the last years, several interesting IDS-based solutions have
been developed to detect insider threats~\cite{insi-3}, 
e.g., searching for confidential 
documents and watermarks in private communications~\cite{encryptedweb}.
 Our approach is directly applicable to these systems. 
 
 Even in scenarios where IDS-based systems may be insufficient,
 \sys can be attractive: \sys is also applicable in more general
 security solutions, for example systems which are based on machine learning
 or anomaly detection of online or offline traffic data, and 
 which can consist of an open-source engine and a possibly confidential
 configuration.
More concretely, advanced methods to detect insider attacks usually consider 
 a wider context when processing network events, therefore requiring to 
 store and query the entire event history \cite{splunk-insider}, 
 and include machine learning algorithms \cite{proactive-insider} (e.g., anomaly detection). 
 Our design can be extended to support these methods, 
 by utilizing the secure storage to save the events history and to
 execute the 
 advanced inspection as an enclaved process that analyzes 
 the history instead of the current traffic. 
 In these cases, our extended system also allows to split the open source 
 inspection engine from the inspection rules, e.g., the big data queries or the machine learning parameters and filters, while
 supporting very general notions of privacy~\cite{privacy2,privacy3,privacy1}. 
 
\subsection{Network Outsourcing}

Another use case arises in the context of network 
outsourcing: We currently witness the trend that 
enterprises wish to outsource their 
cyber security logic or even the entire network administration 
to an external company experienced in this field. 
Translating this scenario to our framework, the enterprise 
(or any of its employees and their PCs) 
constitutes the client, executing the \sys agent, 
and the external 
company operates the \sys server, 
executing the enclaved (and open-source) \sys software.

\subsection{Anti-Terror Intelligence}

An interesting use case also arises in the context of 
governments  gathering intelligence from their own citizens 
in order to fight terror. This use case is highly controversial today and
subject to major ongoing
debates~\cite{KeysUnderDoormats} ,
due to the tradeoff of preserving civil liberties (i.e., the right for privacy) 
 and the government duty to save lives. With the \sys system, 
 governments and citizens can decide and control the level 
 of inspection and privacy infringement applied to the private data. 
The anti-terror intelligence use case may require some 
adaptations to the model. For example, the \sys system can 
be operated by the ISPs that (as part of the \sys software) can 
correlate the traffic with client information. The rules are securely provided 
by the government, but some aggregated information on them can be 
made accessible to the public, or its representatives. The exact matches of 
user traffic, cannot be provided immediately to the user as this might 
prevent the authorities to effectively react to threats, therefore the \sys system 
may make use of a notification delay which is defined a priori.

\section{Discussion and Related Work}\label{sec:discussion}


In an age where more and more resources
and services are outsourced, 
there is an increasing need for solutions
preserving critical security aspects.
Over the last years, this problem
has been discussed particularly intensively
in the context of cloud computing:
Although users of cloud computing infrastructure may
expect their data to remain confidential, today's clouds
are built using a classical hierarchical security model that
aims only to protect the privileged code~(of the cloud
provider) from untrusted code~(the user's virtual machine),
and does nothing to protect user data from access
by privileged code~\cite{haven}.
While for many large-scale computations
today, the use of cloud computing resources
is unavoidable or at least financially 
very attractive, users may not be willing
to trust their cloud provider 
to keep their data confidential.
In fact, the cloud user must trust
not only the hardware on which her or his
data is actually analyzed, but in addition
also  
(i) the provider's software, including privileged software
such as a hypervisor and firmware but also the provider's
full stack of management software;~(ii) the provider's
staff, including system administrators but also those with
physical access to hardware such as cleaners and security
guards;~(iii) the law
enforcement bodies in any jurisdiction where their data
may be replicated, as the Snowden leaks have
revealed~\cite{haven,snowden}.

\begin{table*}[t]
  \centering
\begin{footnotesize}
\begin{tabular}{cc|c|c|c|c|l}
\cline{3-6}
& & \multicolumn{4}{ c| }{\textbf{Architecture}} \\ \cline{3-6}
& & \textbf{Proxy} & \textbf{Client-Side Firewall} & \textbf{BlindBox~\cite{blindbox}} & \textbf{\sys} \\ \cline{1-6}
\multicolumn{1}{ |c  }{\multirow{2}{*}{\textbf{Privacy}} } &
\multicolumn{1}{ |c| }{user exposure} & middlebox & no & no & no~(enclaved) &     \\ \cline{2-6}
\multicolumn{1}{ |c  }{}                        &
\multicolumn{1}{ |c| }{rules exposure} & middlebox & endpoint & middlebox & no~(enclaved)  &     \\ \cline{1-6}
\multicolumn{1}{ |c  }{\multirow{2}{*}{\textbf{Effectiveness}} } &
\multicolumn{1}{ |c| }{inspection guarantee} & yes & no & no & yes &  \\ \cline{2-6}
\multicolumn{1}{ |c  }{}                        &
\multicolumn{1}{ |c| }{supports rules} & any & any & exact match only & any &  \\ \cline{1-6}
\multicolumn{1}{ |c  }{\multirow{2}{*}{\textbf{Overhead}} } &
\multicolumn{1}{ |c| }{computation} & en-\&decryption & none & tokenization \& encryption & decryption &  \\ \cline{2-6}
\multicolumn{1}{ |c  }{}                        &
\multicolumn{1}{ |c| }{communication} & none & none & a stream of tokens & one packet &  \\ \cline{1-6}
\end{tabular}
\end{footnotesize}
\caption{Comparison of performance and security of different IDS architectures.}
\label{tab:tab}
\end{table*}

Recently, researchers have started investigating
whether similar approaches as proposed for cloud
computing may also be applicable
in the context of computer networks, which come
with rather different requirements.
Existing systems are based on proxies and
are vulnerable to a man-in-the-middle attack
on SSL, installing fake certificates at the middlebox~\cite{forged,runa}. 
The middlebox can break the security
of SSL and decrypt the traffic so it can perform the Deep-Packet Inspection~(DPI). 
The removal of the SSL end-to-end security, results in a host
of issues. 
Some proposals allow users to tunnel their traffic to a third
party middlebox provider~\cite{meddle,someone}. 
But these approaches allow the
middlebox owner to inspect/read all traffic. 
An alternative today are distributed firewalls, a client-side
approach to implement intrusion detection/prevention.

A very interesting approach is taken by BlindBox~\cite{blindbox}:
unlike \sys, BlindBox
performs the deep-packet inspection directly on the
\emph{encrypted traffic}. 
In a nutshell, in BlindBox, the endpoint generates a tokenized 
version of the traffic which can be inspected in privacy preserving 
manner at a special server. This tokenized traffic is sent in parallel to the 
origin, increasing the load in the network. 
However, BlindBox requires the user to compute hashes 
of traffic segments and to send them to the inspection box, 
thereby introducing computational overhead to the user PC, as well as 
traffic overhead to the network. Moreover it is dependent 
on the cooperation of the user PC to perform its part 
of the scheme. In addition, BlindBox only supports exact match rules and 
not regular expressions, that are commonly used in security policies.

We believe that our approach nicely
complements these works, and focuses
on a relevant use case, namely traffic inspection. 
An SGX approach
as recently suggested in the context of cloud computing,
could be used to implement our secure server,
and the decoupling of the various
roles identified in our paper.

Table~\ref{tab:tab} summarizes the advantages and disadvantages
of the different architectures: the man-in-the-middle proxy
and the client-side distributed firewalls used today, as well
as BlindBox and \sys. Among these solutions, only BlindBox
and \sys are providing the required privacy guarantees.
The main limitation of BlindBox is arguably its
expressive power, while the main limitation of \sys
is its dependency on a SGX hardware~(although a single box is sufficient).

\section{Conclusion}\label{sec:conclusion}

This paper studied the classic problem of traffic inspection from an
interesting new, 
privacy preserving perspective. In particular,
while today it is commonly believed that 
it is inevitable
that users have to blindly trust the administrator managing the intrusion detection
or prevention system, we in this paper, have questioned this assumption.
In particular, we have shown that
it actually is possible to reduce trust assumptions
in the enterprise network, and presented an intrusion detection
system which is not only privacy preserving regarding
the user traffic but also regarding the rules used
in the IDS/IPS. Interestingly, the proposed \sys
system requires a single secured server; no
modifications of the hardware at the users is required.

In summary, the \sys system features the following properties:
\begin{enumerate}
\item It
decrypts and inspects network traffic in a privacy preserving manner.

\item  It accepts new security rules from administrators and 
applies them to the traffic in 
a secure and privacy-preserving manner: the inspecting device does not
leak any unnecessary information about the user traffic. 

\item  It can be configured with new rules from rule providers 
in a secure and privacy-preserving manner: the inspecting device does not
 leak any unnecessary information about the user traffic. 
\end{enumerate}

We believe that the ideas underlying \sys can be used in
several contexts and beyond the use cases 
discussed in this paper.
In particular, it is not limited to rule-based intrusion detection
systems, but can also be useful in the context of more sophisticated
and offline systems, as they may for example be required
to handle advanced insider threats~\cite{insi-1,insi-2,insi-3}.

Our approach raises several interesting
questions for future research. Obviously,
the performance of our architecture needs to be
evaluated in detail. Moreover, it will be interesting
to explore further the applications of trusted
execution environments in the context of
computer networking and network function virtualization.

{
\def\noopsort#1{} \def\No{\kern-.25em\lower.2ex\hbox{\char'27}}
  \def\no#1{\relax} \def\http#1{{\\{\small\tt
  http://www-litp.ibp.fr:80/{$\sim$}#1}}}\def\noopsort#1{}
  \def\No{\kern-.25em\lower.2ex\hbox{\char'27}} \def\no#1{\relax}
  \def\http#1{{\\{\small\tt http://www-litp.ibp.fr:80/{$\sim$}#1}}}

}

\end{document}